\documentclass{ifacconf}
\usepackage{amsmath}
\usepackage{graphicx}      
\usepackage{natbib}        
\usepackage{amssymb}
\usepackage{multirow}
\usepackage{subcaption}
\usepackage{xcolor}
\usepackage{mathtools}
\DeclarePairedDelimiter{\ceil}{\lceil}{\rceil}
\newcommand{\0}{\mbox{\fontencoding{U}\fontfamily{bbold}\selectfont0}}

\begin{document}
\begin{frontmatter}

\title{Closed Loop Reference Optimization for Extrusion Additive Manufacturing} 

\thanks[footnoteinfo]{Research supported by the NCCR Automation, a National Centre of Competence in Research, funded by the Swiss National Science Foundation (grant number 51NF40\_225155) and by the BRIDGE Discovery programme of the Swiss National Science Foundation and Innosuisse (grant number 226437).}

\author[First]{Rawan Hoteit}, 
\author[Second]{Andrea Balestra},
\author[Second]{Nathan Mingard},
\author[First,Second]{Efe C. Balta},
\author[First]{John Lygeros}

\address[First]{Department of Information Technology and Electrical Engineering at ETH Z\"urich, Switzerland (e-mail: \{rhoteit, jlygeros\}@ethz.ch)}
\address[Second]{inspire AG, Z\"urich, Switzerland (e-mail: \{andrea.balestra, nathan.mingard, efe.balta\}@inspire.ch)}

\begin{abstract}                
Various defects occur during material extrusion additive manufacturing processes that degrade the quality of the 3D printed parts and lead to significant material waste. This motivates feedback control of the extrusion process to mitigate defects and prevent print failure. We propose a linear quadratic regulator (LQR) for closed-loop control with force feedback to provide accurate width tracking of the extruded filament. Furthermore, we propose preemptive optimization of the reference force given to the LQR that accounts for the performance of the LQR and generates the optimal reference for the closed loop extrusion dynamics and machine constraints. Simulation results demonstrate the improved tracking performance and response time. Experiments on a Fused Filament Fabrication 3D printer showcase a root mean square error improvement of $39.57 \%$ compared to tracking the unmodified reference as well as an $83.7\%$ shorter settling time.
\end{abstract}


\end{frontmatter}

\section{Introduction}
Additive manufacturing (AM), or 3D printing, has recently gained considerable traction in a range of industries, from aerospace and automotive engineering to biomedical applications. One such process is Fused Filament Fabrication (FFF), in which plastic is melted and extruded from an extrusion head that is attached to a 3-axis gantry system (\cite{singh2020current}). Various disturbances that occur during the AM process compromise the quality of the print, due to over- and under-extrusion. Extrusion errors result in unintended accumulation or lack of filament at a specific print location in the layer. Moreover, extrusion errors generally lead to considerable post-processing or complete part failure, resulting in significant material waste. 

Over-extrusion is often evident around corners where the extrusion head slows down, causing an accumulation of material at that location. This necessitates modeling the extrusion dynamics (\cite{bellini2004liquefier}) to predict and prevent such defects. Exploiting the interplay between the motion and extrusion dynamics, \cite{kuipers2020framework} generate contour-parallel tool-paths with varying widths via back-pressure compensation. \cite{zhang2024modeling} use flow front boundary equations to model the resulting extrusion height and width, while \cite{Numerical2022Balta} characterize the cross-sectional geometry of deposited material in extrusion AM using numerical models. A simplified linear extrusion model is used by \cite{chesser2019extrusion} to design a feedforward controller that minimizes the tradeoff between printing speed and printing resolution. Moreover, \cite{wu2021accurate} design a feedforward controller based on nonlinear extrusion dynamics to inhibit over- and under-extrusion defects, and present a framework to coordinate motion and extrusion to operate in a region that validates the simplified linear extrusion model. Subsequent work by \cite{wu2023modeling} considers discontinuous extrusion flow rate due to the retraction effects on the filament. \cite{gao2025using} design and validate a nonlinear lead-based feedforward controller after modeling the varying time constant associated with the extrusion model for different printing speeds. Furthermore, a feedforward iterative learning controller is implemented by \cite{bahrami2025optimal} for direct ink writing while modeling the interplay between the extrusion and motion dynamics, using vision-based width estimates.  

Most of the schemes above rely on feedforward control to compensate for over- and under-extrusion.  Feedforward control suffers from open-loop limitations and cannot compensate for sudden defects during printing. Thus, reducing these effects requires in-process monitoring, control, and optimization. In similar AM processes, in-process control is proposed by \cite{shi2018closed} using a PI controller for laser metal deposition printing. Moreover, \cite{rabiei2025extrusion} design a PID to control the extrusion rate for concrete AM using the extrusion pressure measured using a load cell. A model predictive controller is approximated by a neural network by \cite{zomorodi2016extrusion} for various ceramic printing modes. 

According to \cite{moretti2021process}, in-process monitoring and feedback are considered essential in FFF to inhibit extrusion defects or correct them if they were to occur. \cite{fravolini2025data} compute the optimal extrusion feed rate via a Quadratic Programming (QP) formulation that is constrained on linear extrusion dynamics and system constraints. Additionally, \cite{doi:10.1089/3dp.2021.0236} design a PID-based filament transport controller into the printing extruder, while \cite{hornus2020variable} determine varying extrusion width-based paths to minimize extrusion error. Incorporating the linearized dynamics and constraints of the extruder into a model predictive controller is proposed by \cite{xia2020model} based on a vision-based system for extrusion width measurement.

Recent work proposes an in-situ force measurement setup called Force Controlled Printing (FCP). First demonstrated with PID control, FCP is shown to achieve state of the art width control (\cite{guidetti2024force}), and is used in parametric optimization of control parameters (\cite{guidetti2024data}). This work explicitly establishes a relationship between the extrusion force and the resulting filament width on the print bed, and enables the closed-loop control of the geometry of the extruded filament. 

Here we go beyond this state of the art and utilize a linear quadratic regulator (LQR) for extrusion control, based on the FCP setup. We wrap the lower level controller within an optimization scheme that can anticipate the short-comings of the closed loop system in a formulation similar to the reference governor (\cite{kolmanovsky2014reference}). Specifically, the contributions of the paper are as follows: 
\begin{itemize}
    \item The formulation of an optimization problem for reference generation for closed loop extrusion control in additive manufacturing;
    \item The consideration of sim-to-real gap and automatic generation of G-code based on spatiotemporal constraints of the physical system.
\end{itemize}
To validate the contributions, we implement the proposed algorithm in real-time on a hardware setup, achieving a reduction in the tracking error and quicker response time. The design and implementation of such an architecture is novel for extrusion-based additive manufacturing applications.

The paper is organized as follows: Section \ref{Section: Architecture} presents the dynamics and complete control architecture including the optimization formulation. Then, the simulation and experimental results can be seen in Section \ref{Section: Test}; finally, the paper is concluded in Section \ref{Section: Conclusion}.

\section{System Architecture} \label{Section: Architecture}
\subsection{Dynamics} \label{Subsection: Dynamics}
\begin{figure}[h]
    \centering
    \includegraphics[width=0.6\linewidth]{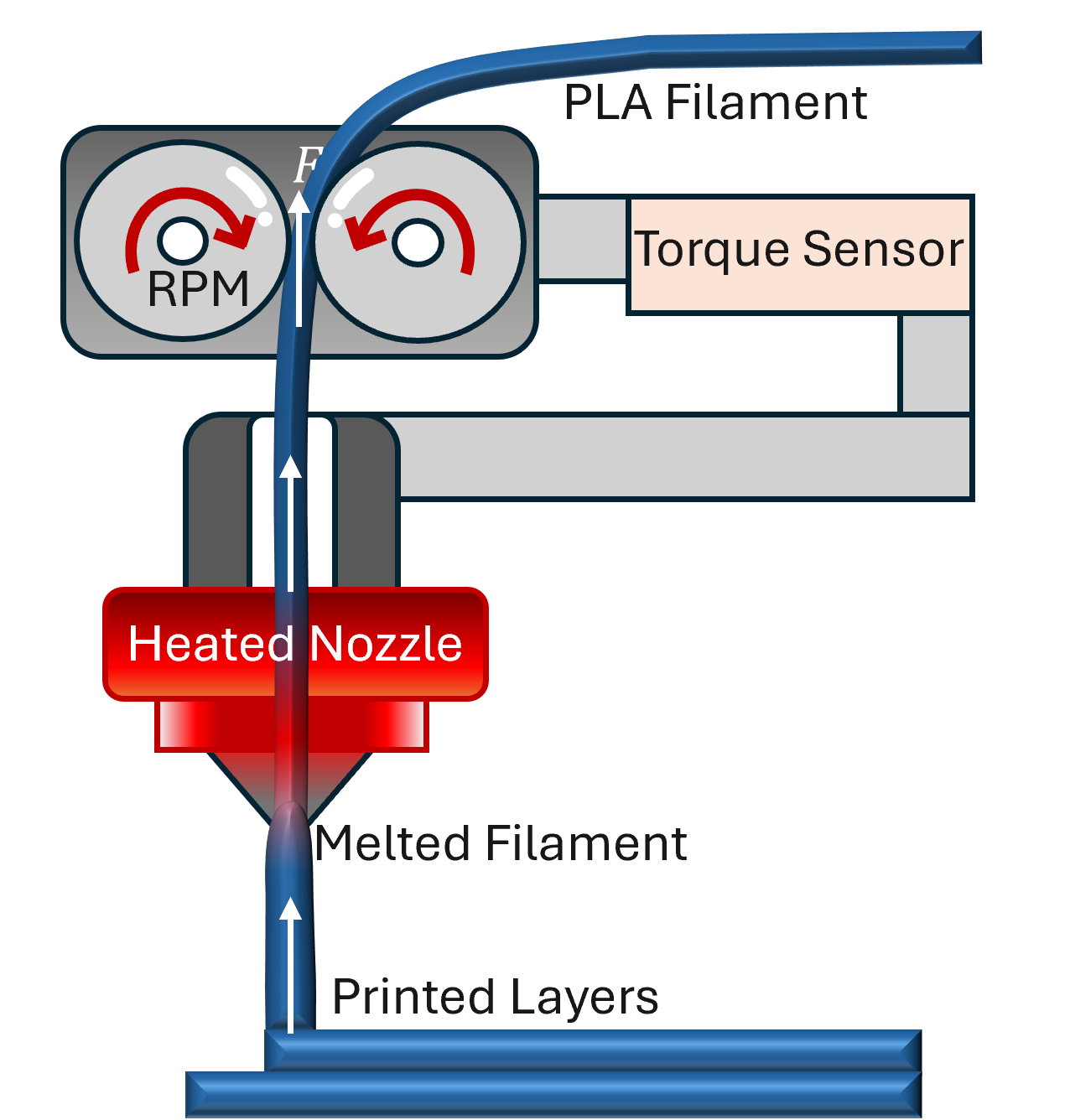}
    \caption{Fused Filament Fabrication (FFF) Process}
    \label{fig: FFF}
\end{figure}

The extrusion head in FFF (Fig. \ref{fig: FFF}) comprises a motor attached to two rollers  that feed a polymer-based filament into a heated nozzle. The nozzle melts the filament to deposit a plastic bead. The extruder then moves in space and forms lines of specific widths onto a heated printing bed. \cite{guidetti2024force} demonstrate that the relationship between the reactive force experienced by the rollers and the line width is linear; this allows one to indirectly measure the line width by measuring the reactive force and inverting the linear relationship. The reaction force can be measured using the force sensor designed by \cite{guidetti2024force}, as illustrated in Fig. \ref{fig: FFF}. \\
The extrusion dynamics of the 3D printer are assumed to follow a discrete time state space representation:
\begin{equation} \label{eq: sys}
\begin{aligned}
   & {x}_{k+1} = A{x}_k+Bu_k, \\
    & {F}_k = C{x}_k,
    \end{aligned}
\end{equation}

where ${x_k} \in \mathbb{R}^n$ is the system state obtained through system identification of the model linking the RPM to the measured force, $u_k$ is the input corresponding to the motor RPM, ${F_k}$ is the measured force and $A, B$ and $C$ are the discrete time system matrices of appropriate dimensions. An LQR controller can be designed based on \eqref{eq: sys} to provide output-reference tracking. The input at time instance $k$ is then given by:
\begin{equation}
    u_k = -K_{LQR}({x}_k-x_{ss,k})+u_{ss,k}.
\end{equation} 
The steady state values of the state $x_{ss,k}$ and of the input $u_{ss,k}$ are computed as a function of the reference force $r_F$ to be tracked:
\begin{equation}
    \begin{bmatrix}
I_{n}-A ~&~ -\!B \\
C ~&~ 0
\end{bmatrix}\begin{bmatrix}
x_{ss,k}\\
u_{ss,k}
\end{bmatrix}= 
\begin{bmatrix}
\0_{n}\\r_{F,k},
\end{bmatrix}\\
\end{equation}
where $I_n$ is the identity matrix with dimensions $n \times n$ and $\0_n$ is the vector of all-zeroes in $\mathbb{R}^n$. $K_{LQR} = (R+B^\top PB)^{-1}B^\top PA$ is the LQR gain computed using the solution of the discrete time Algebraic Riccati Equation:
\begin{equation}
    P = A^\top PA-(A^\top PB)(R+B^\top PB)^{-1}(B^\top PA)+Q,
\end{equation}
such that $Q$ and $R$ are the symmetric weight matrices associated with $x_k$ and $u_k$ in the LQR cost function:
\begin{equation}\label{eq:LQR Cost}
    J = \sum_{t=1}^T(x_t^\top Q x_t+u_t^\top R u_t).
\end{equation}
\subsection{Optimization Formulation}
\begin{figure}[h]
    \centering
    \includegraphics[width=0.9\linewidth]{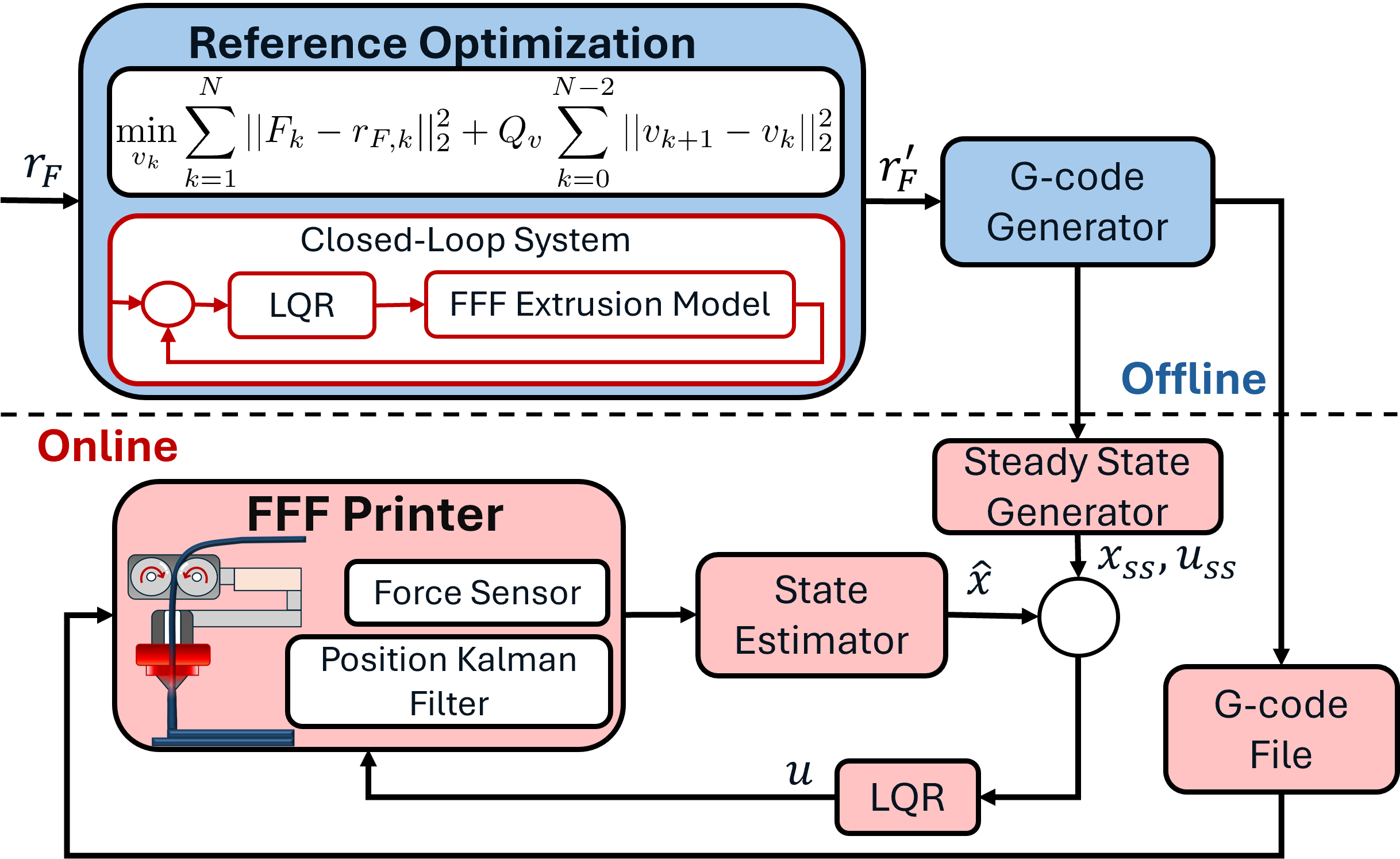}
    \caption{Control Architecture}
    \label{fig: Block Diagram}
\end{figure}

The translation from space into the time domain is a function of the dynamics and printing speed. The desired width profile in space occurs at specific time instances based on the predicted motion and speed of the extrusion head. This allows for the generation of $r_F$ as a function of the time step using the discrete sampling time. Thereby, we generate a sequence of $N$ reference forces $r_F$. However, if we give these directly to the LQR controller to track, the resulting force will differ from the desired one. Though LQR provides enhanced force tracking compared to PID controllers, it is still purely feedback based, and suffers from performance limitations imposed by the limitations of the closed loop system, including long settling times. To provide preview information to the LQR controller, a QP problem is formulated to obtain the optimal reference force given to the closed-loop system (Fig. \ref{fig: Block Diagram}).

This is accomplished by solving the optimization problem:

\begin{equation} \label{eq: QP}
\begin{aligned}
\min_{v_{k}} & \sum_{k=1}^{N}||F_{k}-r_{F,k}||_{2}^2+Q_v\sum_{k=0}^{N-2}||v_{k+1}-v_k||_2^2\\
 s.t & \quad {x}_{k+1} = A{{x}_{k}}+Bu_k \quad \forall \ k = 0,..,N{-}1 \\
& \quad F_{k} = C{x}_{k}  \quad \forall \ k = 0,..,N\\
& \quad \begin{bmatrix}
I_{n}-A ~&~ -\!B \\
C ~&~ 0
\end{bmatrix}\begin{bmatrix}
x_{ss,k}\\
u_{ss,k}
\end{bmatrix}= 
\begin{bmatrix}
\0_{n}\\r_{F,k} +v_k
\end{bmatrix}\\
& \quad u_k = {-}K_{LQR}({x}{-}x_{ss,k})+u_{ss,k},\quad \forall \ k = 0,..,N{-}1\\
& \quad u_{k} \in \mathcal{U} \quad \forall \ k = 0,..,N{-}1\\
& \quad r_F^\prime \in \mathcal{R} \quad \forall \ k = 0,..,N{-}1\\
& \quad x_{k} \in \mathcal{X}. \quad \forall \ k = 0,..,N
\end{aligned}
\end{equation} 

The input to the QP is a sequence of $N$ reference forces, while the output is a modified sequence $r_F^\prime = r_F+v$ that is given to the LQR. The aim is that the resulting force $F_k$ of the closed-loop system tracks the original desired reference $r_F$ at time step $k$, compensating for the closed-loop performance of the extrusion system, while constraining the resulting force reference within thresholds set by the machine. In this work, we set the system limits to reflect the region where system identification was performed. The formulation also minimizes the variation of $v_k$ to ensure smooth modified references between consecutive step sizes through an additional term in the objective function weighted by $Q_v$.

The optimal reference forces are found offline for the LQR to track in run-time. We assume the Kalman filter converges on a faster time scale and the filter dynamics can be omitted. The QP~\eqref{eq: QP} accounts for the closed-loop system dynamics and is solved offline as opposed to receding horizon formulations. Additionally, \eqref{eq: QP} deviates from reference governors since we are solving the problem offline and not enforcing terminal invariance constraints in this preliminary application.

\subsection{G-code Generator}

Applying the inputs computed for the discrete time model \eqref{eq: sys} directly to the system can cause real-time implementation issues, due to the communication delays in the G-code transmission. We account for this delay by imposing a zero-order hold on $v_k$ over multiple time steps to obtain the optimal reference before generating the appropriate G-code. The number of steps over which $v_k$ is held constant is denoted by the hold length $N_h$, which is lower-bounded by the execution frequency. This intrinsically translates into fixing a value of $v$ for each $j = 1, N_h .., \ceil{N/N_h}$ or setting equality constraints in the optimization problem in \eqref{eq: QP}:
\begin{equation}
    v_k = c_j, \quad \forall \ k = 1,..,N, \ j = 1, N_h,..,\ceil{N/N_h}
\end{equation}
where $c_j$ is a constant and $\ceil{.}$ is the ceiling function. 

 After solving the QP and obtaining $r_F^\prime$, the G-code is generated using the motion-based mapping between space and time. The steady state values given to the LQR are recomputed online as $r_F^\prime$ changes. Thus, this methodology considers the sim-to-real gap in terms of the communication pipeline and optimizes for the approximate discretized solution offline, making it executable on a real system.
\section{Testing and Validation} \label{Section: Test}
\subsection{System Identification}
\begin{figure}[ht]
    \centering
    \includegraphics[width=0.7\linewidth]{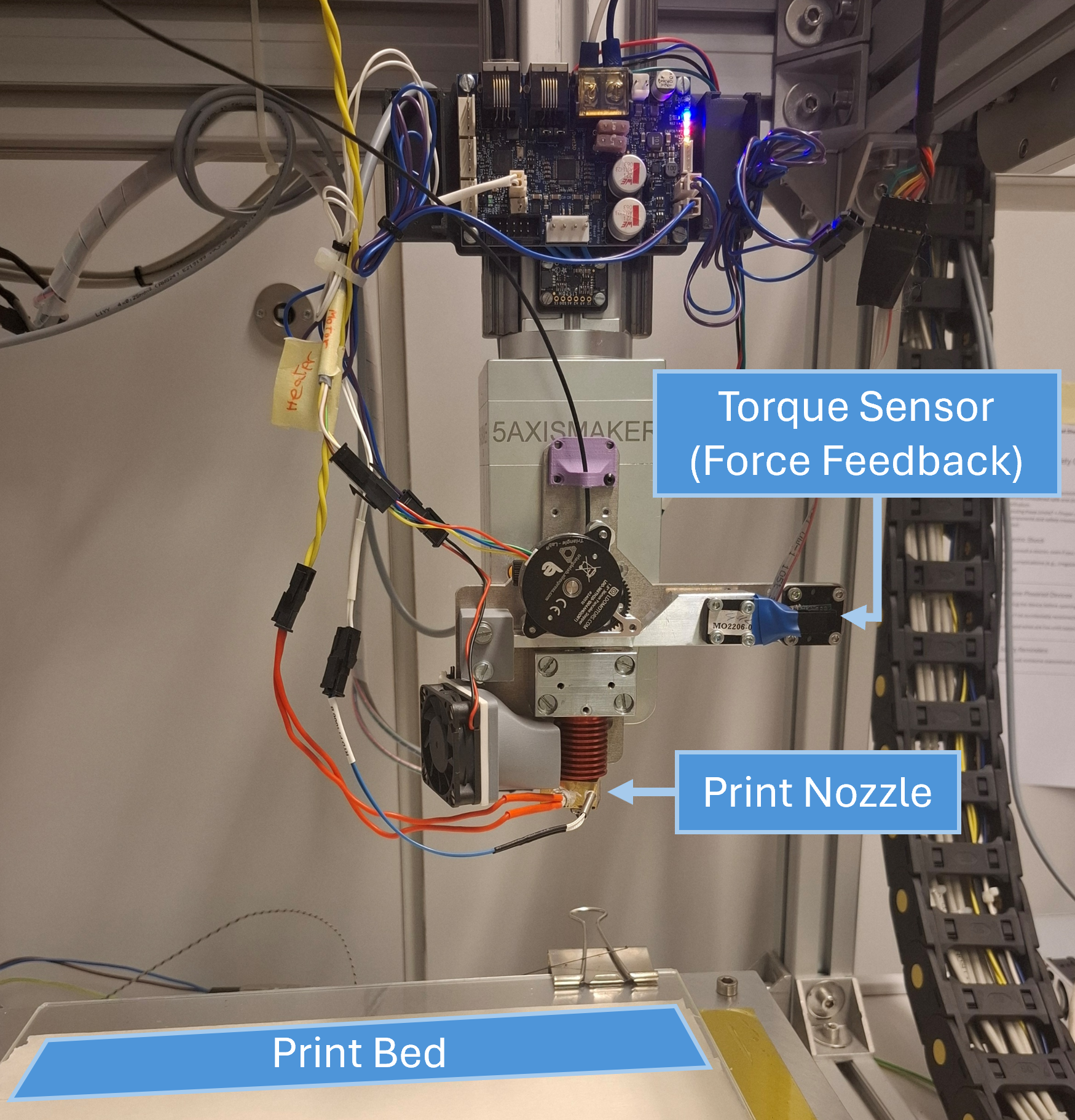}
    \caption{Experimental Setup}
    \label{fig: Setup}
\end{figure}

The FFF extrusion head is attached to a 5-axis gantry which can move along 3 axes ($x,y,z$) and the printing head can rotate along two axes; however, we constrain the motion along those two axes and solely implement planar printing throughout this paper. The printing bed, extrusion head/nozzle and corresponding force sensor can also be seen in Fig. \ref{fig: Setup}. Note that the force value -as given from the sensor- has an inherent negative value.
 
By performing experiments on the system described in Section \ref{Subsection: Dynamics} and seen in Fig. \ref{fig: Setup}, data was collected to identify the extrusion system dynamics \eqref{eq: sys}. Fixing $\tau_s=0.01$ s,  $n=3$, and using the N4SID method by \cite{van1994n4sid} gave rise to the matrices:
\begin{align*}
   & A = 
    \begin{bmatrix}
        1.00603451& 0.01305934& 0.0357625\\
        0.00625487& 1.01087517& 0.0192488\\
       -0.33079381&-0.7101034& 0.5566053\\
    \end{bmatrix},\\
    & B =
    \begin{bmatrix}
        0.00008626\\
-0.00008873\\
0.0047217\\
    \end{bmatrix},\\
    & C = 
    \begin{bmatrix}
        -27.8759035&0.22352502&-0.04037422
    \end{bmatrix}.
\end{align*}

Fixing
\begin{equation}
    Q = \begin{bmatrix}
        1656.2& 0.0& 0.0\\
        0.0& 8.9& 0.0\\
        0.0& 0.0& 1.6\\
    \end{bmatrix}, \quad R = 0.00995,
\end{equation}
then gave rise to the LQR controller with corresponding gain $K_{LQR} = \begin{bmatrix}
    323.8591 &-113.4687 &  23.2255
\end{bmatrix}$.
\subsection{Validation in Simulation}
\begin{figure}
\centering
\begin{subfigure}{\columnwidth}
    \includegraphics[width = \linewidth]{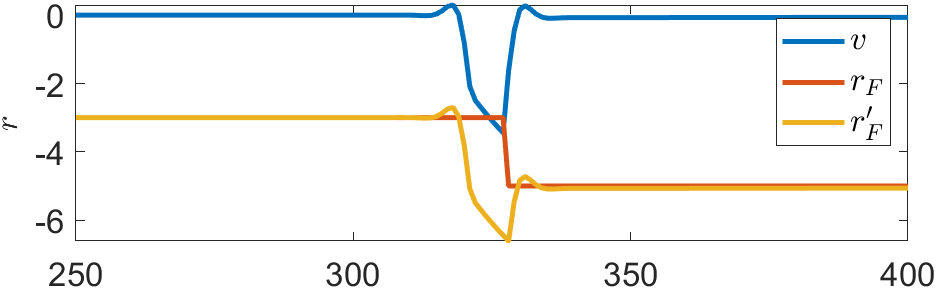}
\end{subfigure}
\hfill
\begin{subfigure}{\columnwidth}
    \includegraphics[width = \linewidth]{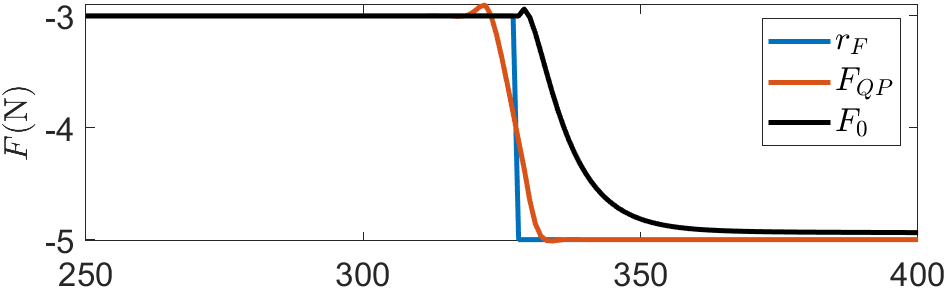}
\end{subfigure}
\hfill
\begin{subfigure}{\columnwidth}
    \includegraphics[width = \linewidth]{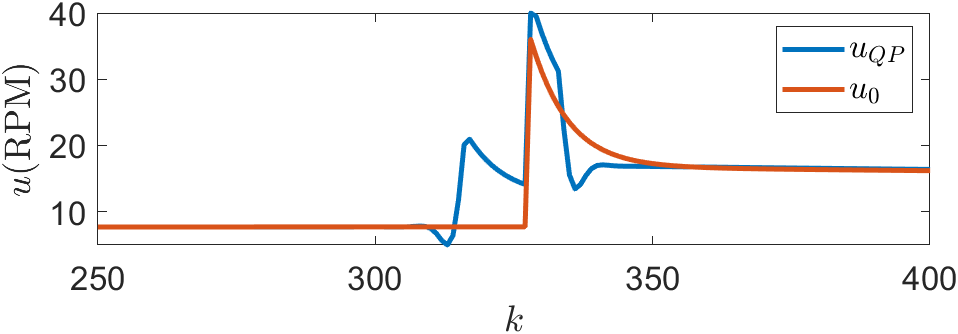}
\end{subfigure}
\caption{Optimization of Step Reference in Simulation- Top: Original and Optimized Reference - Middle: Simulated Force - Bottom: Extrusion RPM Input}
\label{fig: Step_fine}
\end{figure}

In the first test, the reference force to be tracked by the LQR is given by a step function and optimized for over the full horizon via \eqref{eq: QP} (Fig. \ref{fig: Step_fine}). The resulting force is shown in the second plot of Fig. \ref{fig: Step_fine}, having better tracking of $r_F$ in comparison to the original force without optimizing the reference ($F_0$). The new reference $r_F^\prime$ is specifically modified near the time steps when the step occurs, for which the resulting extrusion force is $F_{QP}$. This is to account for the predicted behavior of the closed-loop system which has a large settling time whilst tracking the original reference.  To quantify the improvement, consider the root mean squared error (RMSE)
\begin{equation}
    RMSE(F,r_F) = \sqrt{\frac{1}{N}\sum_{k=1}^N(F_k-r_{F,k})^2},
\end{equation}
and the $5\%$ settling time of the response $t_{5\%}$, defined as the time taken by the system to reach and remain within $5\%$ of the final desired reference value. 
Commanding the extrusion system to track $r_F^\prime$ leads to a RMSE of $0.0637$N that is $69.8 \%$ smaller than the RMSE of $0.211$N obtained when commanding the system to track $r_F$. Correspondingly, $t_{5\%}$ decreases from $0.185$s when commanding the extrusion system to track $r_F$ to $0.035$s when commanding the system to track $r_F^\prime$, a reduction of $81.08\%$.
The corresponding RPM input is shown in Fig. \ref{fig: Step_fine}. The input, $u_{QP}$, needed to track $r_F^\prime$ is significantly larger and is applied earlier than the input for $r_F$, $u_{0}$. This is due to the preview provided by the QP that compensates for the slow response of the system. 

\begin{figure}[h]
    \centering
    \begin{subfigure}{\columnwidth}
    \includegraphics[width = \linewidth]{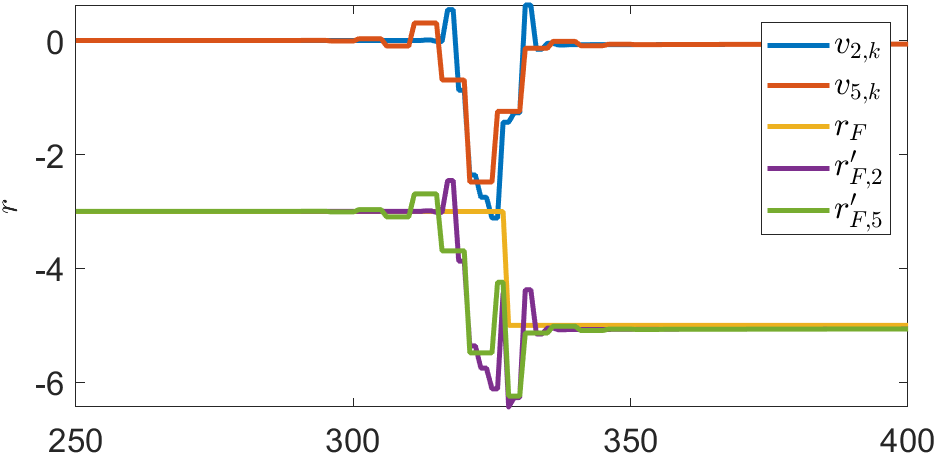}
    \end{subfigure}
    \hfill
    \begin{subfigure}{\columnwidth} 
    \includegraphics[width = \linewidth]{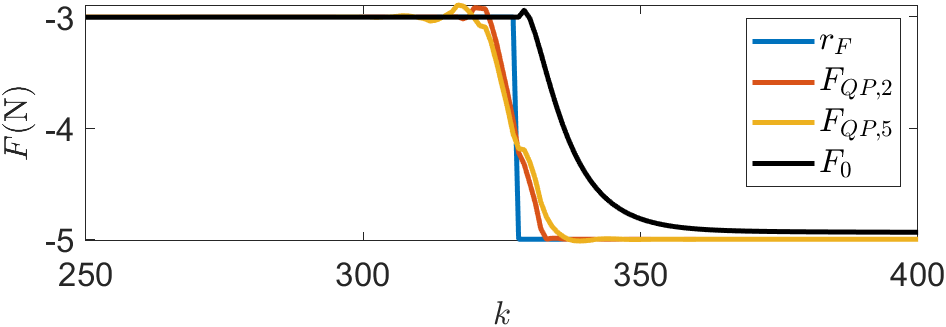}
    \end{subfigure}
    \caption{Optimization of Step Reference in Simulation with $N_h = \{2,5\}$ - Top: Original and Discretized Optimal Force Reference - Bottom: Extrusion Force for Different Discretization Lengths}
    \label{fig: sim_force_coarse}
\end{figure}

 To analyze the impact of various discretizations on the reference optimization, we consider $v_k$ having a hold length $N_h= 2$ in $v_{2,k}$, and $N_h = 5$ in $v_{5,k}$. The resulting optimized reference for a step function from $-3$N to $-5$N for the different discretizations and the corresponding forces are presented in Fig. \ref{fig: sim_force_coarse}; a summary of the results can be found in Table 1. $F_{QP,5}$, the force corresponding to $v_{5,k}$ and $F_{QP,2}$ (resulting from applying $v_{2,k}$) also lead to performance improvements in terms of RMSE and settling time. Compared to applying $r_F$, albeit larger than the one obtained with $N_h=1$, having $N_h =2$ or $N_h =5$ leads to a RMSE that is $7.06\%$ or $32.81\%$ respectively larger than that for $F_{QP}$. Increasing the hold length also leads to slightly higher settling time in comparison to the original QP. However, we emphasize that the advantages of applying a zero-order hold are not apparent in simulation, but come into play in the experimental validation.

\begin{table}[h]
    \centering
    \begin{tabular}{|c|c|c|c|c|}
    \hline
         \textbf{Metric} & $r_F$ & \textbf{$N_h = 1$} & \textbf{$N_h = 2$} & \textbf{$N_h = 5$} \\
         \hline
         RMSE (N) & 0.211 & 0.0637& 0.0682 & 0.0846 \\
         \hline
         $t_{5\%}$ (s)& 0.185 & 0.035 & 0.045 & 0.055\\
         \hline
    \end{tabular}
    \caption{Simulation Performance Comparison}
    \label{tab: sim performance}
\end{table}

\subsection{Validation in Experiments}
 The experimental validation is implemented on the setup seen in Fig. \ref{fig: Setup}. A Kalman filter is employed to estimate the position of the extrusion head; this is important for the synchronization of the force command with the corresponding desired position of the extrusion head along the printing bed. The synchronization between motion and extrusion is implemented via G-code commands given to the 3D printer for trajectory tracking and extrusion control based on the printing speed of the extrusion head. The system is run in real time using a custom ROS2 interface that implements the state estimators, and a Duet controller board that controls the motion and heating. The ROS2 system is responsible for the computation of $x_{ss,k}$ and $u_{ss,k}$, as well as parsing the G-code.

\begin{figure}
    \centering
    \begin{subfigure}{\columnwidth}
        \includegraphics[width = \linewidth]{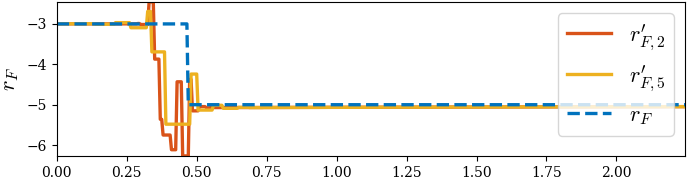}
    \end{subfigure}
    \begin{subfigure}{\columnwidth}
        \includegraphics[width = \linewidth]{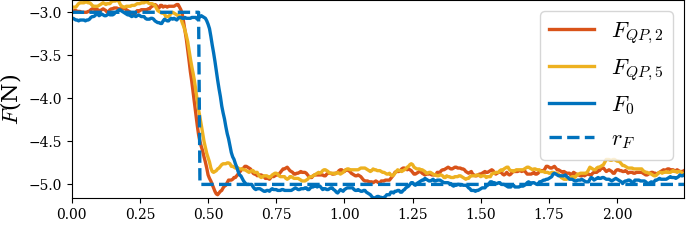}
    \end{subfigure}
    \begin{subfigure}{\columnwidth}
        \includegraphics[width = \linewidth]{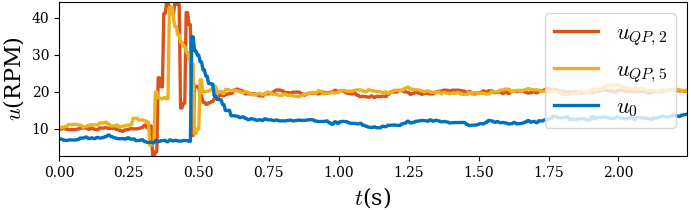}
    \end{subfigure}
    \caption{Step Response on the Experimental Setup - Top: Force  References - Middle: Force Measurements - Bottom: Extrusion RPM Input}
    \label{fig: exp step}
\end{figure}

The step response in Fig. \ref{fig: Step_fine} for which the $r_F$ increases in value from $3$N to $5$N (given by a step from $-3$N to $-5$N per the chosen coordinate system) is tested with $N_h = 2$ and $N_h = 5$; the corresponding optimal references $r_{F,2}^\prime$ and $r_{F,5}^\prime$ can be seen in Fig. \ref{fig: exp step} with the respective performance of the extrusion force. The input RPM applied on the filament is presented in Fig. \ref{fig:  exp step} showcasing a similar trend to that in Fig. \ref{fig: Step_fine} with $u_{QP}$ acting prior to the step time and having a higher value than that  of $u_{0}$. This leads to enhanced performance in comparison to the case without reference optimization. Furthermore, increasing $N_h$ from $2$ to $5$ results in a smoother RPM signal which provides slightly lower RMSE and slower response.

\begin{figure}
    \centering
\includegraphics[width=\linewidth]{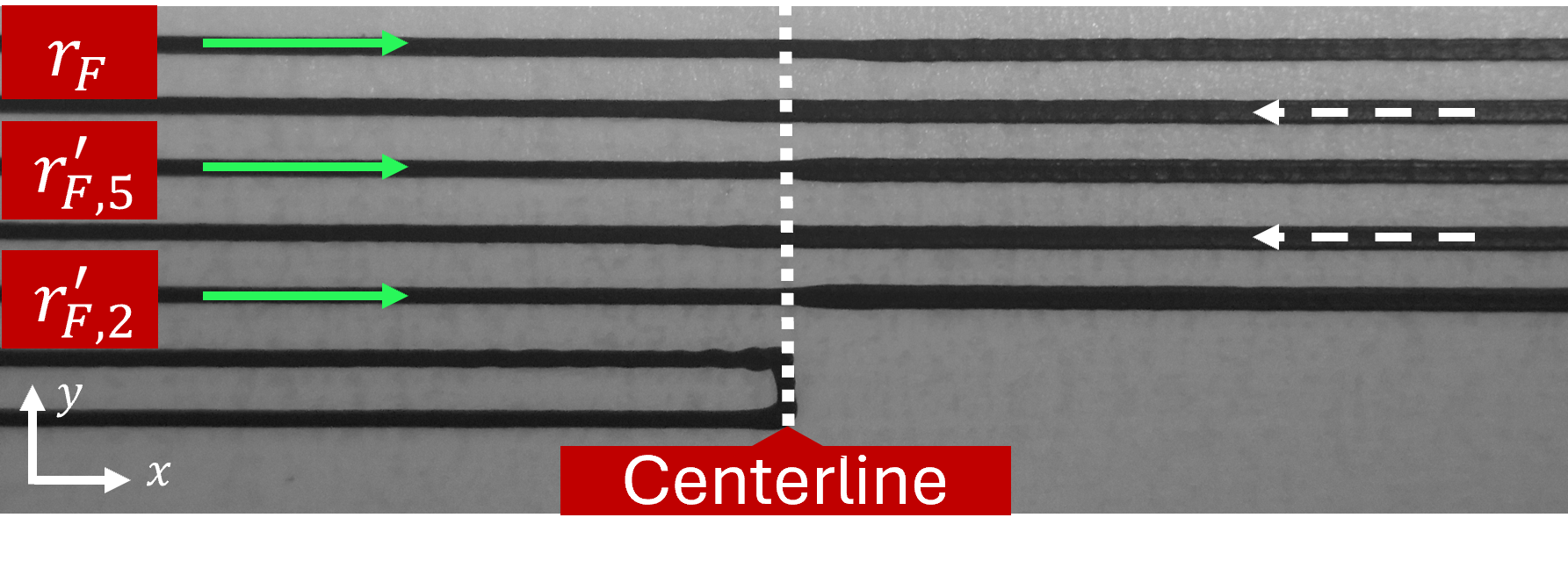}
    \caption{Step Print Lines}
    \label{fig: step print}
\end{figure}

The improved response time can also be seen in the resulting print shown in Fig. \ref{fig: step print}. The step from $-3 \text{N}$ to $-5\text{N}$ occurs halfway through the horizontal line as the print head moves from left to right as indicated in Fig. \ref{fig: step print}. The white arrows indicate lines where the force is returned back to $-3 \text{N}$ before progressing into the next method. The lines were printed from bottom to top, starting with $r_{F,2}^\prime$ then $r_{F,5}^\prime$ and finally $r_F$ which shows the performance of the LQR. The slight overshoot in $F_{QP,2}$ that can be seen in the middle plot of Fig. \ref{fig:  exp step} is not evident in Fig. \ref{fig: step print} and does not seem to compromise the print quality.

Though the performance significantly improves, the difference in results between the simulations and experimental validation is due to the model mismatch in the simplifications of the extrusion dynamics during the system identification process. 

A rheology-based analysis of the filament extrusion dynamics reveals its nonlinearity with respect to the extrusion force (\cite{bellini2004liquefier}). However, linearization is necessary to design the LQR controller for real-time feedback and control actions during the print. Moreover, linearization leads to a tractable optimization problem when optimizing the reference offline. While increased model fidelity improves reference tracking performance, it introduces increased computational requirements.

\begin{table}[h]
    \centering
    \begin{tabular}{|c|c|c|c||c|c|}
    \hline
         \textbf{Metric} & \multicolumn{1}{|c|}{\textbf{LQR}} & \multicolumn{2}{|c||}{\textbf{LQR+QP}}& \multicolumn{2}{c|}{\textbf{Improvement}}\\
         \hline 
          &  &$N_h = 2$ & {$N_h = 5$}&$N_h = 2$ & {$N_h = 5$} \\
         \hline
         RMSE (N) & 0.278 &  0.173 & \textbf{0.168}&  37.77 \% & \textbf{39.57 \%} \\
         \hline
         $t_{5\%}$(s) & 0.135 &  \textbf{0.022} &  0.039& \textbf{83.70 \%}& 71.11 \% \\
         \hline
    \end{tabular}
    \caption{Experimental Performance Metrics}
    \label{tab: performance}
\end{table}

As the optimization problem is run offline there is no run-time computational overhead or concerns for real time implementation; rather the main effort lies in coordinating the extrusion reference with specific spatial positions along the print bed via G-code. 
\begin{figure}[h]
    \centering
\includegraphics[width=\linewidth]{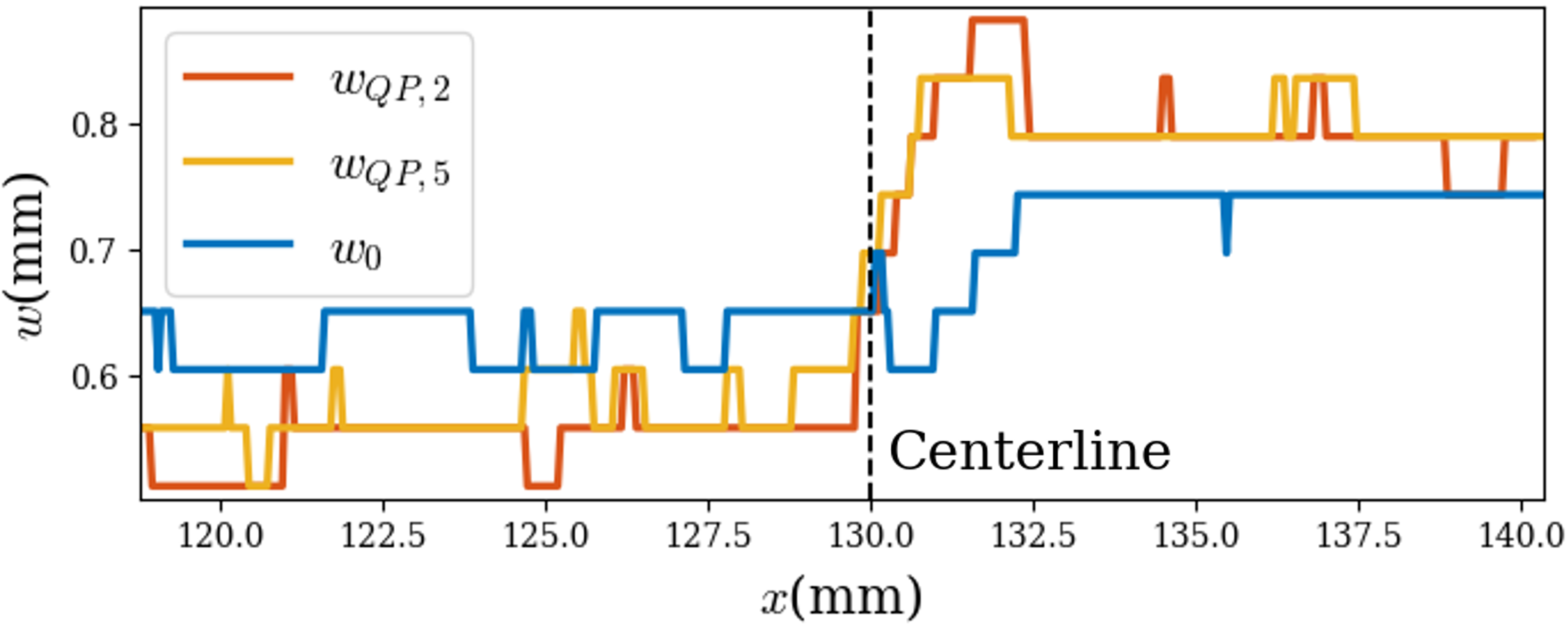}
    \caption{Linewidths from Modified and Original References Extracted from Fig. \ref{fig: step print}}
    \label{fig: width}
\end{figure}
 Through analyzing captured images of the print lines in Fig. \ref{fig: step print}, the respective linewidth $w_0$ can be obtained in Fig. \ref{fig: width} upon applying LQR without modifying the reference force; the figure also presents the width obtained for different hold lengths ($w_{QP,2}$ for $N_h = 2$, and $w_{QP,5}$ for $N_h = 5$). Evidently from Fig. \ref{fig: width} and Fig. \ref{fig: exp step}, the LQR reacts slowly to the change in extrusion, leading to a gradual increase in linewidth. Experimentally, the shortcoming of the reactive LQR can be seen as the thickness of the extruded lines changing after the desired change-point at the center, in contrast to tracking the modified inputs which present an almost instantaneous change in width near the midpoint of the line.
 
 Finally, tuning $Q$ and $R$ in~\eqref{eq:LQR Cost} affects the resulting performance of the closed-loop system. However, this effect is attenuated via reference optimization, which ensures enhanced tracking based on the cost function and constraints in~\eqref{eq: QP} regardless of the underlying controller tuning.
\section{Conclusion} \label{Section: Conclusion}
We present an optimization formulation of the force references to be tracked by FFF printers that compensates for the closed-loop performance of the printer. The formulation is implemented offline taking into consideration the behavior of the system in the presence of an LQR. Simulation results reveal performance improvement of up to $69.81\%$ in RMSE. Experimentally, an improvement in the tracking error by $39.57\%$ and response time by $83.7\%$ is obtained, though model mismatch leads to the deviation of the experimental results from those obtained via simulation. Most importantly, there is no additional run-time computational overhead and the proposed QP can be solved efficiently offline. Future work will investigate nonlinear extrusion behavior and nonlinear real-time control design. Future work also involves implementing the reference optimization online, in the presence of different types of controllers and under various printing conditions.


\bibliography{ifacconf}             

@article{hornus2020variable,
  title={Variable-width contouring for additive manufacturing},
  author={Hornus, Samuel and Kuipers, Tim and Devillers, Olivier and Teillaud, Monique and Mart{\'\i}nez, Jon{\`a}s and Glisse, Marc and Lazard, Sylvain and Lefebvre, Sylvain},
  journal={ACM Transactions on Graphics (TOG)},
  volume={39},
  number={4},
  pages={131--1},
  year={2020},
  publisher={ACM New York, NY, USA}
}

@article{shi2018closed,
  title={Closed-loop control of variable width deposition in laser metal deposition},
  author={Shi, Tuo and Lu, Bingheng and Shen, Ting and Zhang, Rui and Shi, Shihong and Fu, Geyan},
  journal={The International Journal of Advanced Manufacturing Technology},
  volume={97},
  number={9},
  pages={4167--4178},
  year={2018},
  publisher={Springer}
}

@article{guidetti2024force,
  title={Force controlled printing for material extrusion additive manufacturing},
  author={Guidetti, Xavier and Mingard, Nathan and Cruz-Oliver, Raul and Nagel, Yannick and Rueppel, Marvin and Rupenyan, Alisa and Balta, Efe C and Lygeros, John},
  journal={Additive Manufacturing},
  volume={89},
  pages={104297},
  year={2024},
  publisher={Elsevier}
}

@article{xia2020model,
  title={Model predictive control of layer width in wire arc additive manufacturing},
  author={Xia, Chunyang and Pan, Zengxi and Zhang, Shiyu and Polden, Joseph and Wang, Long and Li, Huijun and Xu, Yanling and Chen, Shanben},
  journal={Journal of Manufacturing Processes},
  volume={58},
  pages={179--186},
  year={2020},
  publisher={Elsevier}
}

@article{moretti2021process,
  title={In-process simulation of the extrusion to support optimisation and real-time monitoring in fused filament fabrication},
  author={Moretti, M and Rossi, A and Senin, N},
  journal={Additive Manufacturing},
  volume={38},
  pages={101817},
  year={2021},
  publisher={Elsevier}
}

@article{chesser2019extrusion,
  title={Extrusion control for high quality printing on Big Area Additive Manufacturing (BAAM) systems},
  author={Chesser, Phillip and Post, Brian and Roschli, Alex and Carnal, Charles and Lind, Randall and Borish, Michael and Love, Lonnie},
  journal={Additive Manufacturing},
  volume={28},
  pages={445--455},
  year={2019},
  publisher={Elsevier}
}

@article{wu2021accurate,
  title={Accurate linear and nonlinear model-based feedforward deposition control for material extrusion additive manufacturing},
  author={Wu, Pinyi and Ramani, Keval S and Okwudire, Chinedum E},
  journal={Additive Manufacturing},
  volume={48},
  pages={102389},
  year={2021},
  publisher={Elsevier}
}

@article{wu2023modeling,
  title={Modeling and feedforward control of filament advancement and retraction in material extrusion additive manufacturing},
  author={Wu, Pinyi and Qian, Chen and Okwudire, Chinedum E},
  journal={Additive Manufacturing},
  volume={78},
  pages={103850},
  year={2023},
  publisher={Elsevier}
}

@article{doi:10.1089/3dp.2021.0236,
author = {Moretti, Michele and Rossi, Arianna},
title = {Closed-Loop Filament Feed Control in Fused Filament Fabrication},
journal = {3D Printing and Additive Manufacturing},
volume = {10},
number = {3},
pages = {500-513},
year = {2023}}

@article{zhang2024modeling,
  title={Modeling of deposition morphology and characteristic dimensions in material extrusion additive manufacturing},
  author={Zhang, Jiamin and Wang, Lilin and Lin, Xin and Huang, Weidong},
  journal={Additive Manufacturing},
  volume={89},
  pages={104306},
  year={2024},
  publisher={Elsevier}
}

@inproceedings{guidetti2024data,
  title={Data-driven extrusion force control tuning for 3D printing},
  author={Guidetti, Xavier and Mukne, Ankita and Rueppel, Marvin and Nagel, Yannick and Balta, Efe C and Lygeros, John},
  booktitle={2024 IEEE 20th International Conference on Automation Science and Engineering (CASE)},
  pages={2262--2267},
  year={2024},
  organization={IEEE}
}

@article{fravolini2025data,
  title={Data-driven reference shaping for optimal fused filament fabrication},
  author={Fravolini, Mario Luca and Rossi, Arianna and Moretti, Michele and Senin, Nicola and Ferrante, Francesco},
  journal={Control Engineering Practice},
  volume={164},
  pages={106511},
  year={2025},
  publisher={Elsevier}
}

@inproceedings{kolmanovsky2014reference,
  title={Reference and command governors: A tutorial on their theory and automotive applications},
  author={Kolmanovsky, Ilya and Garone, Emanuele and Di Cairano, Stefano},
  booktitle={2014 American Control Conference},
  pages={226--241},
  year={2014},
  organization={IEEE}
}

@article{gao2025using,
  title={Using nonlinear lead filtering for real-time accurate extrusion control in large format additive manufacturing},
  author={Gao, Jutang and Wu, Pinyi and Okwudire, Chinedum E and McGee, Wes},
  journal={Additive Manufacturing},
  pages={105005},
  year={2025},
  publisher={Elsevier}
}

@article{kuipers2020framework,
  title={A framework for adaptive width control of dense contour-parallel toolpaths in fused deposition modeling},
  author={Kuipers, Tim and Doubrovski, Eugeni L and Wu, Jun and Wang, Charlie CL},
  journal={Computer-Aided Design},
  volume={128},
  pages={102907},
  year={2020},
  publisher={Elsevier}
}

@article{rabiei2025extrusion,
  title={Extrusion under material uncertainty with pressure-based closed-loop feedback control in robotic concrete additive manufacturing},
  author={Rabiei, Mahsa and Moini, Reza},
  journal={Automation in Construction},
  volume={180},
  pages={106494},
  year={2025},
  publisher={Elsevier}
}

@inproceedings{zomorodi2016extrusion,
  title={Extrusion based additive manufacturing using explicit model predictive control},
  author={Zomorodi, Hesam and Landers, Robert G},
  booktitle={2016 American Control Conference (ACC)},
  pages={1747--1752},
  year={2016},
  organization={IEEE}
}

@inproceedings{bahrami2025optimal,
  title={Optimal Feed-Forward and Iterative Learning Control Framework for Enhanced Precision in Extrusion-Based Additive Manufacturing},
  author={Bahrami, Ali and Watson, Christopher and Tilbury, Dawn and Barton, Kira},
  booktitle={International Manufacturing Science and Engineering Conference},
  volume={89015},
  pages={V001T01A012},
  year={2025},
  organization={American Society of Mechanical Engineers}
}

@article{bellini2004liquefier,
  title = {Liquefier Dynamics in Fused Deposition},
  volume = {126},
  number = {2},
  journal = {Journal of Manufacturing Science and Engineering},
  publisher = {ASME International},
  author = {Bellini,  Anna and G\"{u}\c{c}eri,  Sel\c{c}uk and Bertoldi,  Maurizio},
  year = {2004},
  month = may,
  pages = {237–246}
}

@article{singh2020current,
  title={Current status and future directions of fused filament fabrication},
  author={Singh, Sunpreet and Singh, Gurminder and Prakash, Chander and Ramakrishna, Seeram},
  journal={Journal of Manufacturing Processes},
  volume={55},
  pages={288--306},
  year={2020},
  publisher={Elsevier}
}

@article{Numerical2022Balta,
    author = {Balta, Efe C. and Altınkaynak, Atakan},
    title = {Numerical and experimental analysis of bead cross-sectional geometry in fused filament fabrication},
    journal = {Rapid Prototyping Journal},
    volume = {28},
    number = {10},
    pages = {1882-1894},
    year = {2022},
    month = {05},
    issn = {1355-2546},
    eprint = {https://www.emerald.com/rpj/article-pdf/28/10/1882/2881872/rpj-09-2021-0255.pdf},
}

@article{van1994n4sid,
  title={N4SID: Subspace algorithms for the identification of combined deterministic-stochastic systems},
  author={Van Overschee, Peter and De Moor, Bart},
  journal={Automatica},
  volume={30},
  number={1},
  pages={75--93},
  year={1994},
  publisher={Elsevier}
}
                                                   







\appendix
\end{document}